# Structural relaxation in the hydrogen-bonding liquids N-methylacetamide and water studied by optical Kerr-effect spectroscopy

**David A. Turton and Klaas Wynne**

*Department of Physics, SUPA, University of Strathclyde, Glasgow G4 0NG, Scotland, UK*

Structural relaxation in the peptide model N-methylacetamide (NMA) is studied experimentally by ultrafast optical Kerr-effect spectroscopy over the normal-liquid temperature range and compared to the relaxation measured in water at room temperature. It is seen that in both hydrogen-bonding liquids, β relaxation is present and in each case it is found that this can be described by the Cole-Cole function. For NMA in this temperature range, the α and β relaxations are each found to have an Arrhenius temperature dependence with indistinguishable activation energies. It is known that the variations on the Debye function, including the Cole-Cole function, are unphysical, and we introduce two general modifications: one allows for the initial rise of the function, determined by the librational frequencies, and the second allows the function to be terminated in the α relaxation.

## INTRODUCTION

Collectively referred to as structural relaxation, the lowest-frequency modes observed in the spectra of molecular liquids are the strongly overdamped modes which arise from molecular interactions and which reflect diffusive molecular reorientations involving the formation and breaking of weak intermolecular bonds. The study of relaxation can therefore yield insight into the timescale and degree of intermolecular bonding. In early studies, structural relaxation was often analyzed in terms of simple Debye functions (exponential decays in the time domain). However, improvements in sensitivity and bandwidth have enabled greater accuracy in analyses and the identification of more complex behavior, particularly in associating liquids such as those capable of intermolecular hydrogen bonding as well as in supercooled and glass-forming liquids.

Structural relaxation is bounded at high frequency by the librational modes, which typically occur in the terahertz region, and at low frequency by the α relaxation mode. Additional relaxation modes observed at higher frequencies than α relaxation are known as intermediate, secondary, or β relaxations and their origins are diverse and often less clear. Here we measure the OKE response of the biologically important hydrogen-bonding molecule N-methylacetamide (NMA). Whilst we show that previous suggestions that NMA has an anomalous temperature dependence of hydrogen bonding[1,2] are incorrect, we resolve α and β relaxations, which appear indicative of the degree of hydrogen bonding, and relate this behavior to the relaxation decay measured in water.

Relaxation in liquids has been studied experimentally by a wide range of diffraction, scattering, and absorption techniques. Dielectric relaxation spectroscopy (DRS) has been widely applied, particularly to slower systems such as glass-forming liquids,[3] but also to hydrogen-bonding liquids.[4,5] DRS is effective over many decades of frequency up to about 100 GHz above which terahertz and far-infrared absorption spectroscopy must be used to extend the range.[5,6] Optical Kerr-effect spectroscopy (OKE),[7-10] which we use here, is to some extent a complementary technique offering good sensitivity up to about ten terahertz from the sub-gigahertz region in a single measurement. OKE has previously been used to measure relaxation in many applications including simple liquids,[9,11] ionic liquids,[12,13] solutions of peptides,[8,14] hydrogen-bonding liquids,[2,15-18] and supercooled and glass-forming liquids.[19-22]

Many models have been applied to relaxation in liquids. In the simplest cases, for example rotational relaxation of simple spheroids, relaxation can be modeled as a single Debye function $1/(1+i\omega\tau)$.[23] For rigid molecules of low symmetry, the general case for rotational relaxation measured through four-wave mixing has been shown to be equivalent to the sum of five Debye functions.[23,24] However, since the polarizability tensor typically contains few non-zero elements, and mild asymmetry appears not to have a strong influence on diffusional rates, most molecular relaxation can be described considerably more simply than this. In practice, simple models of relaxation have been applied to a surprisingly wide range of liquids.

However, for more complex systems, and in particular strongly associating liquids, relaxation modes are observed that cannot be fit by a simple Debye function. In these cases, functions such as Cole-Cole,[25,26] Cole-Davidson,[4,22,27] or Havriliak-Negami[28] (in the frequency domain) or Kohlrausch-Williams-Watts[21,29,30] (KWW) (in the time domain) have been employed. These functions were originally introduced as empirical fitting models but have since been shown to be consistent with theoretical models of diffusion.[31,32] They can each be seen as a broadening of the Debye function as would arise from a superposition of Debye processes due to heterogeneity of a liquid. Alternatively, the KWW function has been interpreted as a slowing of relaxation due to a transition from single molecule to cooperative motion.[30] Although the KWW and Cole-Cole functions are quite similar, for glass-forming liquids the KWW is widely found to be a good model of the α relaxation whereas the Cole-Cole function is more often applied to β relaxations.

Even in the study of simple liquids,[33,34] for example symmetrical tops, where orientational relaxation is expected to yield a single exponential decay, double exponential decays are observed. The origin of the faster intermediate or β relaxation is generally assigned to interaction- or collision-induced effects.[33-35] For unstructured liquids where the intermediate decay appears exponential, it has been interpreted in the context of Kubo theory as arising from "motional narrowing".[10,18,36] Even in mono-atomic liquids, where orientational relaxation is absent, an (interaction-induced) relaxation signal is still observed.[35]

In supercooled liquids, β relaxations[37] have been studied widely and assigned to the correlation of short-range translational motions within a transient "cage" of neighboring particles which is clearly similar to the interaction-induced behavior observed in simple liquids. In this picture α relaxation is then the complete decay of correlations as the cage breaks up. The α and β relaxations are often readily separated due to the emergence of different temperature dependences below the melting point whereas these might merge and become indistinguishable at higher temperatures.[26] This type of behavior has been investigated within the framework of mode coupling theory (MCT).[19,21,38]

Like water, NMA is a hydrogen-bonding liquid of great biophysical importance which has been studied intensively both experimentally and theoretically.[1,2,39,40] NMA is essentially a single peptide linkage, terminated by a pair of methyl groups, which adopts the trans conformation overwhelmingly favored in proteins. It is therefore a useful elementary protein-model system. The dipole moment of the HN–CO moiety enables hydrogen bonding and the formation of chains[41-43] and it has recently been suggested that the presence of this higher structure in the pure liquid results in unusual properties including a non-Arrhenius temperature dependence of the structural relaxation.[1,2]

## EXPERIMENT

Our basic OKE setup has been described previously.[8] For this experiment, 800-nm-wavelength pulses with an energy of 8 nJ at a repetition rate of 76 MHz were generated by a Coherent *Mira-SEED* oscillator. After pre-compensation, for group-velocity dispersion in the setup, in a *homosil* prism pair, the beam was split into (90%) pump and (10%) probe beams, which were co-focused by a 10-cm-focal-length achromatic lens into the sample contained in a 2-mm-pathlength quartz cuvette. The auto-correlation function measured at the sample position in a quartz plate, corresponded to a 24-fs (FWHM) sech$^2$ pulse duration. The induced change in polarization of the probe beam was analyzed by the combination of a quarter-wave plate, a Wollaston prism, and an amplified balanced photodiode detector. The variable pump-probe time delay was introduced by a single 250-mm optical delay line with a resolution of 500 nm (3.3 fs). For the relaxation measurements, the step size at delays up to 0.5 ps was 50 fs and thereafter logarithmically increasing to give a total of approximately 250 points. For each trace an average of at least 16 complete scans was taken and before analysis the signals were resampled (by linear interpolation) to a linear timescale. Additional data were taken at 5-fs resolution to measure the librational modes at room temperature.

An accurate estimate of the baseline is essential to analyze the decay function at longer times, and for each scan, 10 ps of baseline signal was measured before time-zero. The baseline throughout the measurement must also be flat, and for this reason, both pump and probe beams were mechanically chopped at rates of about 5 kHz in the ratio of 5:7 with lock-in demodulation at the difference frequency. The double chopper attenuates the signal considerably, but minimizes any spurious signals due to pump light reaching the detector via scattering from the sample and Fresnel reflections at the sample windows. Chopping the probe can result in large transient out-of-balance signals when using balanced detection and to minimize this, each chopper was placed at the focus of a 1:1 ($f$ = 16 cm) telescope. In this configuration the pulse duration after pre-compression increased to ~35 fs.

An anaerobic 2-mm fused quartz cuvette (Starna Optiglass) was rinsed with filtered methanol then dried under vacuum. NMA as supplied by Sigma-Aldrich at >99% purity was centrifugally filtered (0.1 $\mu m$ Millipore) then added to the cuvette. The cuvette was evacuated to approximately 1.5 torr and the temperature increased with gentle agitation to de-gas the sample. Once the liquid was stable the temperature was increased until gentle refluxing was seen (approximately 80 °C) where it was held for 15 minutes to dry the sample. The cuvette was then sealed.

To control the sample temperature, the cuvette was enclosed in a close-fitting copper block. A pair of low-voltage electrical heaters attached to the sides of the block was driven by a pulse-width-modulated temperature controller with a platinum resistance sensor attached directly to the cuvette, close to the illuminated area, to provide temperature feedback. Calibration confirmed the accuracy to be within the 0.1 K precision of the controller.

## RESULTS AND DISCUSSION

NMA is a normal liquid between approximately 300 K and 480 K. The OKE signal was measured at nine temperatures in this range as shown, displayed on logarithmic axes, in Fig 1. The time-domain signal represents the convolution of the auto-correlation with the derivative over time of the two-point time-correlation function of the anisotropic part of the polarizability tensor. The deconvoluted Fourier transform of this signal is equivalent to the depolarized Raman spectrum corrected by the Bose-Einstein population factor.[34,44,45] At the lower temperatures, several cycles of oscillation due to a superposition of librational modes can be seen up to 1 ps. After this, the decay is monotonic through the intermediate region. Then at the longest times, α relaxation becomes apparent.

Mode-coupling theory (MCT) describes the correlation of density-density fluctuations and is particularly appropriate for liquids below the melting point giving insight into the dynamics as the critical temperature, related to the glass temperature, is approached. However the MCT framework has also been applied to liquids above the melting point suggesting that MCT is a more general model of liquid behavior.[46,47] The practical function derived from MCT[48]

$$S(t) = \left(pt^{-z} + dt^{b-1}\right)\exp\left(-t/\tau_\alpha\right) \quad (1)$$

has been found to be a good model of the OKE decay when applied to organic liquids in the normal-liquid to moderately supercooled temperature region as well as ionic liquids, and liquid crystals.[12,49] Here the power law $pt^{-z}$ describes the β relaxation and the exponential term accounts for the α relaxation. The term $dt^{b-1}$ is the "von Schweidler power law" added to represent the "transition" from β to α relaxation. For more complex decays, the number of power-law terms is increased. A crucial aspect of the MCT model is that $z$ and $b$ are temperature-independent and the temperature-dependence of the decay arises from the change in the amplitudes, $p$ and $d$. This model, with three power laws, was applied to the data but a satisfactory fit was not achieved. Specifically, the fitted value for $b$ increased with temperature, giving values greater than one, which appears to be physically meaningless. An alternative model was therefore sought.

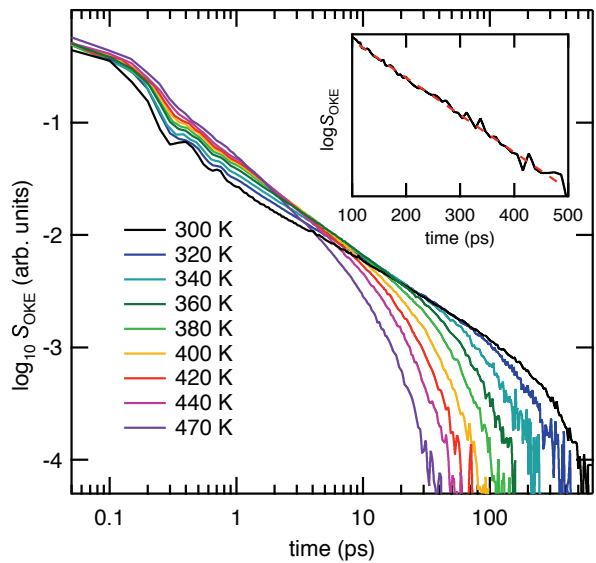

Fig 1. (Color online) Log-log plot of the time-domain OKE signal decay for NMA. All signals have been normalized to the signal peak at zero delay. Inset is detail of the decay at 300 K on a log-linear scale, showing that after 100 ps it follows a simple exponential decay (dashed line).

At lower temperatures the intermediate part of the decay is clearly non-exponential, so multiple exponential models were discarded in favor of KWW and Havriliak-Negami functions. The time-domain KWW function is a "stretched" exponential

$$S_{KWW}(t) = \exp\left(-(t/\tau)^\alpha\right), \quad (2)$$

and the Cole-Cole, Cole-Davidson, and Debye functions are special cases (with reduced parameters) of the Havriliak-Negami function

$$S_{HN}(\omega) = \frac{1}{\left(1 + (i\omega\tau)^\beta\right)^\alpha}. \quad (3)$$

The Cole-Cole function results when $\alpha = 1$,

$$S_{CC}(\omega) = \frac{1}{1 + (i\omega\tau)^\beta}, \quad (4)$$

the Cole-Davidson when $\beta = 1$, and the Debye when $\alpha = \beta = 1$. The Fourier transform of the Cole-Davidson function can be written as

$$S_{\text{CD}}(t) = \sqrt{2\pi} \frac{t^{\beta-1} e^{-t/\tau}}{\Gamma(\beta)\tau^{\beta}}, \quad (5)$$

where it can be seen that the MCT model, expressed as Eq. (1), is also a reduced form of the double-Havriliak-Negami function equivalent to the sum of two Cole-Davidson functions constrained by a common exponential term and with $\beta_1 = 1-z$ and $\beta_2 = b$. For comparison, the time derivative of the KWW as fitted to OKE data can be written as

$$\frac{dS_{\text{KWW}}(t)}{dt} = \beta \frac{t^{\beta-1} e^{-(t/\tau)^{\beta}}}{\tau^{\beta}}. \quad (6)$$

This expression is not equivalent to a Havriliak-Negami function but behaves, in many cases, very similarly to the Cole-Cole function[50]. Neither the KWW or the Cole-Cole function has an analytical Fourier transform.

With respect to the Debye function (exponential decay), the effect of each exponent is to broaden the function in the frequency domain. In certain MCT studies,[47] a relaxation decay has been modeled as a (nearly) logarithmic decay ($z \approx 1$) and we note that both the Cole-Cole and Cole-Davidson functions give a logarithmic decay in the limit of the exponent approaching zero. The Cole-Cole function has been applied previously to relaxation in complex liquids.[25,26,32,51,52]

Given the similarities of these functions, and the complexity of the decay measured for NMA, it is probably impossible to define a unique model. However, the simplest fit that described the relaxation over the measured temperature range was found to be the sum of a Cole-Cole function, for the intermediate decay, and a Debye function accounting for the α relaxation at the longest delay times. However when the librational period before 1 ps was analyzed, it was found that an additional fast overdamped response was present and this appears close to a temperature-independent Debye function. No practical time-domain form of the Cole-Cole function exists, so the fit function was generated entirely in the frequency domain and then numerically transformed to the time domain for fitting to the data.

With respect to the Debye function, the Cole-Cole is broadened (symmetrically on a logarithmic frequency scale) which slows the decay on long timescales (similar to the decay of the KWW function). A limitation of frequency-domain models employing multiple Havriliak-Negami modes (including Cole-Cole, Cole-Davidson, and Debye modes) is that the total relaxation process is modeled as a sum of these modes – as if these modes are completely independent of each other. However, if the Cole-Cole (or KWW) function is used to represent β relaxation, the slow tail will tend to become slower than the α relaxation which would be unphysical.

The solution to this is provided by studies of β relaxations in glass-forming liquids which show that relaxation, to some extent, consists of sequential processes; specifically, β relaxation gives way to, and is then terminated by, the complete structural α relaxation. In the MCT model of Eq. (1), this is taken into account by multiplying each power law with the exponential decay representing the α relaxation. We have therefore introduced a modification to the Cole-Cole function (see Appendix) by taking the time-domain product with the α relaxation function (although performed in the frequency domain). We refer to this as the α termination. This modification allows the relaxation signal to evolve into a pure Debye function and this improves the fit at the longest times.

In an analogous process, the relaxation initially evolves from the anisotropic state created from the superposition of librational fluctuations driven by the envelope of the pump pulse. The relaxation function must therefore include a rise time that will be related to the frequency of the librational modes. This effectively imposes an upper boundary on the relaxation rate (and spectrum) in the same way that the α termination imposes a lower boundary. Although the Debye function has been used very widely, it is unphysical in this respect as it rises instantaneously at $t = 0$ and, in the frequency domain, decays too slowly at high frequency. The Cole-Cole function can be considered worse in this respect due to the broadening, and, for this reason, stochastic models have been favored. This limitation of the Debye function is well-known, and in time-domain OKE studies[11,16,53] it has been overcome empirically by subtracting a fast exponential decay representing the inertial rise time. This is most conveniently achieved by taking the product of the relaxation with the function

$$1 - \exp(-\gamma_{\text{LIB}} t) \quad (7)$$

where $\gamma_{\text{LIB}}$ is the initial rate of the rise determined by the librational frequencies.

For a frequency-domain function $S(\omega)$ (where $S$ is e.g. a Debye or Cole-Cole function) this inertial response can be included in the frequency domain (see Appendix) as

$$S'(\omega) = S(\omega) - S(\omega + i\gamma_{\text{LIB}}) \quad (8)$$

and in the fit, the α relaxation was modeled by this expression with $S$ a Debye function.

The β relaxations are then modeled by modifying the Cole-Cole function to

$$S'(\omega) = S_{\text{CC}}(\omega + i/\tau_{\alpha}) - S_{\text{CC}}(\omega + i/\tau_{\alpha} + i\gamma_{LIB}), \quad (9)$$

which incorporates the α termination. Written in full this is

$$S'(\omega) = \left( \frac{1}{1 + (i\omega\tau + \tau/t_{\alpha})^{\beta}} - \frac{1}{1 + (i\omega\tau + \tau/t_{\alpha} + \gamma_{\text{LIB}}\tau)^{\beta}} \right). \quad (10)$$

The fast and intermediate decays were modeled by this function.

The initial part of the relaxation decay has strong librational modes superimposed, so a reasonable model of these is essential for an accurate fit to the relaxation modes. For OKE studies the Brownian oscillator has been shown to be an accurate model of librations[8,44] and can be written as

$$S_B(\omega) = \omega_0^2 / (\omega_0^2 - \omega(\omega + i\gamma)). \quad (11)$$

Here $\omega_0$ and $\gamma$ are angular frequencies representing the undamped oscillator frequency and the damping rate respectively. Critical damping appears at $\gamma \cong 2\omega_0$ and in the limit of $\gamma \gg \omega_0$ the response reproduces the Debye function. A high-resolution spectrum was taken at 300 K (Fig 2) allowing fitting of the libration modes up to almost 30 THz (1000 cm$^{-1}$).

The fit function was created in the frequency domain starting with an assumed laser pulse temporal profile of $I(t) = \text{sech}^2((t-\Delta t)/a)$. The frequency-domain autocorrelation of this (normalized by area) is

$$AC(\omega) = a^2 \pi \omega^2 \text{cosech}^2(a\pi\omega/2)\exp(i\omega\Delta t). \quad (12)$$

From this, the convoluted fit function was formed by taking its product with the sum of the three relaxation modes, nine librational modes, and an amplitude parameter representing the instantaneous (hyperpolarizability) response. The Fourier transform of the convolution was then fitted to the OKE data by a Levenberg-Marquardt non-linear least-squares algorithm. For the underdamped modes, the initial values for the fit were taken from Reference 39. These and the parameters returned from the fit are shown in Table 1.

| mode | $A$ (a.) | $\gamma$ (THz) | $\omega_0$ (THz) | $\omega_0$ (ref 39) |
|------|----------|----------------|------------------|---------------------|
| $f_1$ | 521 | 3.22 | 2.30 | 2.31 |
| $f_2$ | 92.1 | 1.55 | 2.60 | 2.85 |
| $f_3$ | 135 | 1.94 | 3.30 | 3.63 |
| $f_4$ | 5.05 | 1.00 | 5.92 | 5.64 |
| $f_5$ | 7.12 | 1.10 | 8.81 | 8.57 |
| $f_6$ | 4.56 | 0.31 | 13.12 | 13.10 |
| $f_7$ | 4.27 | 1.33 | 18.09 | 18.26 |
| $f_8$ | 4.98 | 0.31 | 18.78 | 18.83 |
| $f_9$ | 3.02 | 0.56 | 26.42 | 26.47 |

Table 1 Parameters returned by the fit to the 300-K data for NMA shown in Fig 2, for the modes fit by nine ($f_{1-9}$) Brownian oscillators, Eq. (11). Values obtained from Raman spectra of crystalline NMA at 25 °C in Reference 39 are given for comparison.

The exponential term, including $\Delta t$, in Eq. (12) allows for the offset in time between it and the data, to be corrected. Although the peak around zero time is predominantly due to the hyperpolarizability response, there is a significant contribution from the nuclear response and neglecting this delay results in the spectrum becoming negative at higher frequencies. Three parameters in the fit set the amplitude, delay, and width of the autocorrelation. For the data measured at 300 K, these were returned as 0.95 and -2.45 fs (with respect to the peak of the data) with a width that corresponds to a pulse duration of 24 fs (FWHM). We find this process gives greater accuracy than experimental methods of measuring the autocorrelation.

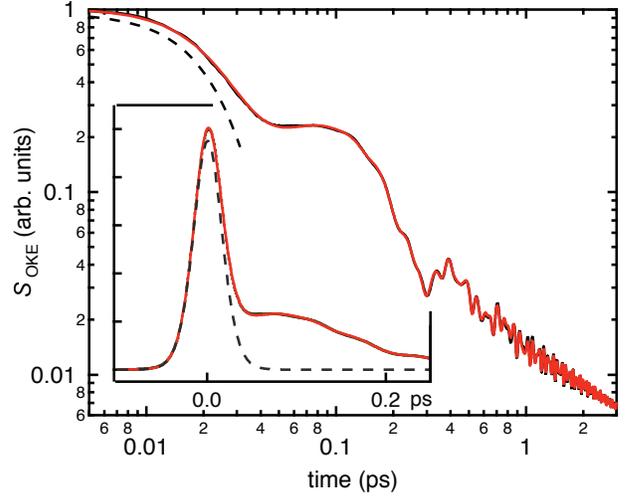

Fig 2. (Color online) High-resolution OKE data for NMA at 300 K shown up to 2.5 ps on a log-log plot and (inset) up to 0.25 ps on a linear plot. The fit is mostly indistinguishable from the data. The (fitted) sech$^2$-autocorrrelation function is shown in both plots as a dashed curve. The fit parameters are given in Table 1.

The temperature-dependent data and their fits are shown in Fig 3. At a 50-fs step size, only the librational modes below 10 THz contribute to the spectrum. In the liquid phase three of these modes and the fast relaxation mode are overlapped and appear in the data as a broad shoulder which cannot be resolved. By reference to the Raman spectrum of the crystalline phase,[39] this band can be assigned to two low-frequency intramolecular modes (expected to be only weakly temperature dependent) and two hydrogen bonding modes. With increasing temperature this region softens and shifts slightly to lower frequency although the large degree of overlap means that this cannot be made quantifiable and furthermore makes the fit to the fast relaxation mode unreliable.

| $T$ (K) | $A(\beta_1)$ (a.u.) | $\beta(\beta_1)$ | $\tau(\beta_1)$ (ps) | $A(\alpha)$ (a.u.) | $\tau(\alpha)$ (ps) |
|---------|---------------------|------------------|----------------------|--------------------|--------------------|
| 300 | 30.1 | 0.58 | 27.9 | 32.7 | 183 |
| 320 | 28.0 | 0.58 | 12.9 | 29.9 | 105 |
| 340 | 28.8 | 0.57 | 9.93 | 23.7 | 60.5 |
| 360 | 28.3 | 0.59 | 6.83 | 19.6 | 38.2 |
| 380 | 26.7 | 0.6 | 4.51 | 17.4 | 25.5 |
| 400 | 25.8 | 0.62 | 3.02 | 15.0 | 17.6 |
| 420 | 23.0 | 0.65 | 1.86 | 13.4 | 12.1 |
| 440 | 20.9 | 0.7 | 1.50 | 12.1 | 8.72 |
| 470 | 18.5 | 0.72 | 1.38 | 9.02 | 6.21 |

Table 2. Parameters for the fits shown in Fig 3. $A(\beta_1)$, $\beta(\beta_1)$, and $\tau(\beta_1)$ refer to the intermediate Cole-Cole relaxation; $A(\alpha)$ and $\tau(\alpha)$ the, Debye, $\alpha$ relaxation. The fast relaxation was held fixed with $\tau(\beta_2)=0.3$ ps and $\beta(\beta_2)=0.94$ and the inertial rise fixed at $\gamma = 14$ THz. The librational modes cannot be fit quantitatively.

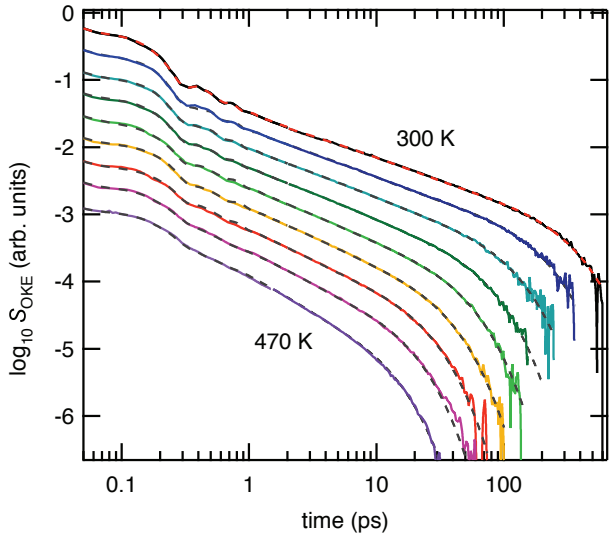

Fig 3. (Color online) Temperature-dependent OKE data for NMA and the fits (dashed lines) as described in the text. All traces except that at 300 K are vertically offset. The temperatures and relaxation-mode parameters are given in Table 2.

At this resolution, the fit is only weakly sensitive to the inertial rise rate $\gamma$, and the value of 14 THz given by the higher resolution fit was applied as a fixed parameter for the temperature-dependent data where it gave satisfactory results throughout. This frequency is naturally higher than the measured librational frequencies because it represents a rise time rather than a full period of oscillation, hence $\gamma_{LIB} \approx \langle \omega_0 \rangle / 2\pi$. The influence of this modification to the Cole-Cole function, in the frequency domain, is shown in Fig 4.

After minimizing the sum of squared residuals, the quality of the fit was judged by eye in a logarithmic plot which, covering several orders of magnitude in both amplitude and time, clearly shows when a simple model is successful. Fig 4 also shows the relationship of the three relaxation functions in the fit to the data for NMA at 360 K. The short time Cole-Cole function ($\beta_2$) with an exponent of 0.94 is close to an exponential decay. It was observed that this fast decay was essentially independent of temperature and, in order to make the fitting more robust, was held constant. The librational modes which can be assigned to hydrogen-bonded intermolecular modes are underdamped but there is considerable overlap in this region in the liquid phase making quantitative fitting impossible. These modes were found to broaden slightly and shift marginally to lower frequency with increasing temperature. With these caveats, the model gave an accurate fit to the dataset over the whole temperature range. These fits are shown in Fig 3 as dashed curves and the fit parameters are shown in Table 2, Fig 5, and Fig 6.

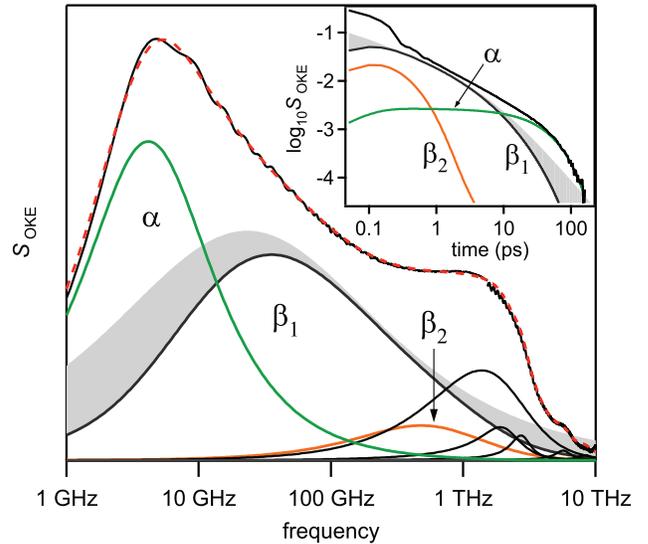

Fig 4. Fit to the 360-K data for NMA shown in the frequency domain. The three relaxation modes are shown as heavy lines accompanied by the librational modes which all peak above 1 THz. The shaded area shows the difference in the Cole-Cole function $\beta_1$ due to the modification to include the rise time and the $\alpha$ termination. The inset compares the three fitted relaxation modes to the data in the time domain.

The OKE experiment measures the derivative of the two-point correlation function of the anisotropic part of the polarizability tensor. For the fundamental ($\alpha$) relaxation, this is generally assumed to be due to orientational diffusion, *i.e.*, the diffusion of a single molecule under the influence of the friction of the surrounding molecules. The Stokes-Einstein-Debye (SED) expression relates the single molecule relaxation rate $D$, for rotational or translational relaxation, to the (macroscopic) shear viscosity $\eta$, and the effective molecular radius $R$:

$$D_{trans} = \frac{k_B T}{6\pi\eta R}, \quad D_{rot} = \frac{k_B T}{8\pi\eta R^3}. \quad (13)$$

These expressions are derived for the Brownian motion of a spherical particle immersed in a viscous homogeneous fluid, but have been successfully applied to many real liquids of small and moderately sized molecules and hold true for spheroids which yield single-exponential dynamics. For pure rotations, the rate of the decay due to the second rank polarizability tensor is proportional to $6D$ and hence the single molecule rotational Debye time-constant (proportional to $1/6D_{rot}$) is given by

$$\tau_r = \frac{V_{eff}\eta}{k_B T} + \tau_r^0, \quad (14)$$

expressed now as a function of the effective molecular volume $V_{eff}$ and where $\tau_r^0$ is the zero-viscosity intercept. OKE measures $\tau_c$, the collective rotational Debye time-constant and this is related through $\tau_c = g_2\tau_r/j_2$ where $g_2$ is introduced as a static pair correlation term and $j_2$ allows for dynamic angular momentum correlations.[9] If these expressions hold true, then α relaxation as measured by OKE should exhibit the same temperature dependence as shear viscosity. Viscosity typically obeys the Arrhenius equation $\eta \propto \exp(E_a/k_B T)$ with activation energy $E_a$. Therefore, if the SED relation holds for this system, the time constant measured by OKE should also have an Arrhenius temperature dependence described by

$$\tau_c(T) = \frac{A}{T}\exp\left(\frac{E_a}{k_B T}\right). \quad (15)$$

This is found to be the rule for most liquids at temperatures well above the glass transition.

The temperature dependence of the time constants for the Debye α relaxation $\tau(\alpha)$, and the Cole-Cole β relaxation $\tau(\beta_1)$ obtained from the fits are shown in Fig 6 along with the results of fitting by Eq. (15). Over this temperature range (we have not succeeded in significantly supercooling NMA), normal Arrhenius behavior is seen in both time constants although there is significantly more scatter in the data for $\tau(\beta_1)$, which is presumed to be due to the overlap of this mode with the higher frequency modes. The measured activation energy $(E_a/k_B)$ for the α relaxation is $(2470 \pm 40)$ K ($E_a = 0.2$ eV). A line representing this activation energy is also drawn through the data for $\tau(\beta_1)$ which demonstrates that the activation energies are indistinguishable within the scatter in $\tau(\beta_1)$.

In previous OKE studies,[1,2] it has been suggested that NMA exhibits a non-Arrhenius temperature dependence of the α relaxation due to an anomalous temperature dependence of the hydrogen-bonded chain length. However, in those cases a simple double-exponential model was fitted to the decay, which was measured over a fixed range of only 25 ps. This range is clearly too short to resolve the α relaxation at lower temperatures and the effect of this is that, as Fig 1 shows, at lower temperature it would be the β relaxation modes that contribute to the fit more than the α relaxation. Studies of near-infrared absorption in NMA[54] are consistent with a continuous distribution of chain lengths and there appear to be no anomalous relaxation dynamics.

A rheological measurement[2] of shear viscosity over the temperature range of approximately 350 K to 420 K also observes a normal Arrhenius temperature dependence with an activation energy of 1500 K. This is significantly lower than our measured value, but there is considerable scatter in the viscosity data at higher temperature. Low-frequency DRS measurements of the α relaxation taken from 305 K to 360 K put the value higher at ~3200 K.[42] However, viscosity data[55] taken more recently over the range of 293 K to 318 K yield an activation energy of 2540 K which is in good agreement with our value.

The Arrhenius dependence implies a coupling between shear viscosity and the collective relaxation measured by OKE spectroscopy. Eq. (14) then allows a value for the effective molecular volume to be calculated. From the literature values for viscosities[55] with the fit to the relaxation data, a value of 180 Å$^3$ can be calculated. This is 40% higher than the 128 Å$^3$ calculated simply from the liquid density[56] (at the melting point) of 0.946 g cm$^{-3}$. but lower than the crystalline unit cell volume, measured by neutron diffraction, of approximately 460 Å$^3$.[57]

Simple liquids often display Debye relaxation modes but structured liquids like NMA and water are likely to be inhomogeneous. The higher structure of water is much debated, but there is a strong consensus that a range of cluster sizes exist and that an individual water molecule experiences a fluctuating range of environments with differing degrees of hydrogen bonding. Similarly, NMA forms chains in the pure liquid with a temperature-dependent range of lengths. The effect of the exponent in the Cole-Cole function is to broaden the Debye function (symmetrically on a logarithmic frequency axis) and this can be interpreted as inhomogeneity in the hydrogen-bonding environment. In Fig 5 the Cole-Cole exponent for the β relaxation can be seen to rise somewhat as temperature increases suggesting a reduction in the inhomogeneity consistent with a reduction in the extent of hydrogen bonding. However the change is small and again is subject to the uncertainty in the higher frequency modes.

For water, it is well known that hydrogen bonding results in structure which, to a first order, is tetrahedral although the extended structure is still debated. The model of relaxation is informative on the extent of structuring and for water has been studied in both time and frequency domains.[17,18,21,51,58] In time-domain studies the double-exponential model has generally been used, but with poor agreement in the measured parameters. A superior fit in the time domain has been shown for the KWW function[21] and, in the frequency domain, for the Cole-Cole function.[51]

Despite the extensive study of water, the origin of the measured decay signals is still unclear. Dielectric and infrared spectroscopies measure the same properties as OKE and Raman spectroscopies but with essential differences: the former interact through the total dipole moment and measure the first rank ($\ell = 1$) orientational relaxation rate whereas OKE and Raman scattering experiments measure the polarizability–polarizability correlation function and therefore the second rank ($\ell = 2$) orientational relaxation rate which appears three-times faster.[59] In addition, water has a near-isotropic polarizability tensor[60] therefore the intermolecular contribution arises predominantly from interaction-induced effects due to translational motions, of hydrogen-bonded water molecules, rather than from rotational motions.[61] The rotational contribution to the OKE relaxation signal is then extremely weak and, for water, the OKE signal is almost complementary to that of DRS.[62] In Ref.[5], the DRS spectrum for water at 298 K is fitted by the sum of two Debye functions with time constants of 8.4 ps for the α relaxation and 1.1 ps for an intermediate β-relaxation mode that is very weak and at the extreme of the frequency range. As the polarizability of water is low, the relaxation can be measured over only a few orders of magnitude in our OKE setup; the decay to 10 ps at room temperature is shown in Fig 7. The relaxational part of this data can be fit by a single Cole-Cole function with $\tau_\beta = 0.61$ ps and $\beta_\beta = 0.86$ (with $\gamma_{LIB} = 12$ THz). This timescale is similar to that of the β relaxation measured in DRS and we draw the conclusion that the relaxation measured in water by OKE is a β relaxation which is accompanied by an unmeasurable α relaxation.

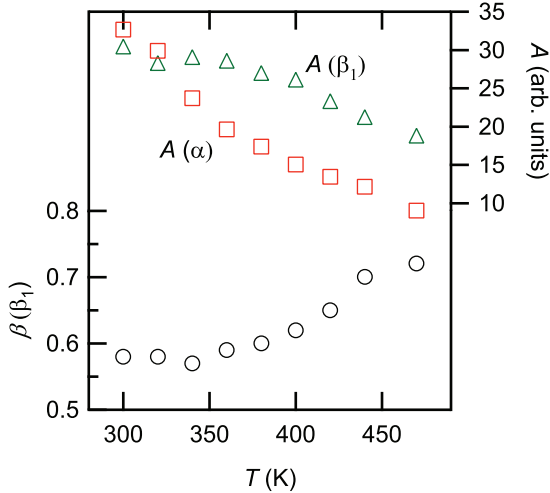

Fig 5. Parameters for the fits shown in Fig 3. $A(\beta_1)$ and $\beta(\beta_1)$ refer to the amplitude and Cole-Cole exponent for the intermediate β relaxation, and $A(\alpha)$ is the amplitude of the, Debye, α relaxation.

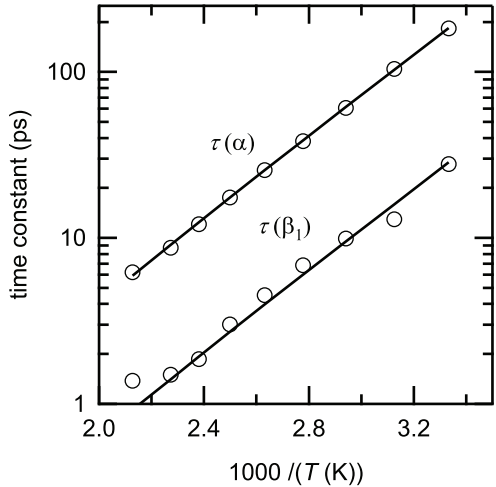

Fig 6. The temperature dependence of the time constants for the α relaxation, $\tau(\alpha)$, and the intermediate β relaxation, $\tau(\beta_1)$. The data for $\tau(\alpha)$ are fitted by an Arrhenius dependency, Eq. (15), while a line of the same activation energy is drawn through the data for $\tau(\beta_1)$.

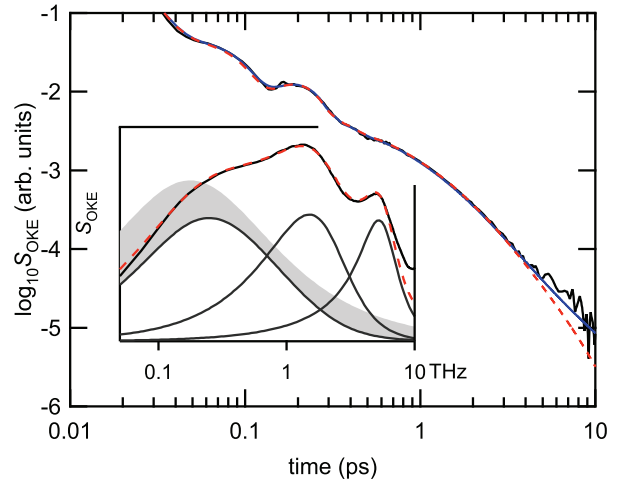

Fig 7. (Color online) OKE signal for water and the fit of the sum of a Cole-Cole function and five Brownian oscillators (solid and dashed lines, where the dashed refers to the fit which includes the α termination). Inset is the frequency-

*domain representation including the fit (dashed) which includes the α termination. The fit has also been decomposed to show the β-relaxation mode and the first two Brownian oscillators, $f_1$ & $f_2$. The shaded part shows the modification to the Cole-Cole function by the inertial rise and α termination. The fit parameters are given in Table 3.*

Consequently, although the α relaxation is not apparent in the OKE data, it must still be present as a termination to the β relaxation. Therefore the data were refitted, with Eq. (10) representing the β relaxation with the value of 2.8 ps (8.4 ps /3) for $\tau_\alpha$. Both these fits to the data are shown in the time domain in Fig 7 where the difference can be seen at times greater than 4 ps. The modified function has parameters of $\tau_\beta$ = 1.0 ps and $\beta_\beta$ = 0.84. The modified fit is poorer here but this may be a shortcoming of the data. The signal-to-noise ratio of the data is clearly inadequate to prove the difference in these models and whether or not the α relaxation is actually detectable; future work with an amplified-laser setup will attempt to resolve these issues.

| mode | $A$ (a. u.) | $\beta$ | $\tau$ (ps) | |
|---|---|---|---|---|
| β | 9.78 | 0.86 | 0.61 | |
| β' | 12.5 | 0.86 | 1.01 | |
| | $A$ (a. u.) | $\gamma$ (THz) | $\omega_0$ (THz) | $\omega_0$ (cm$^{-1}$) |
| $f_1$ | 219 | 3.62 | 2.19 | 73 |
| $f_1$' | 236 | 3.81 | 2.21 | 74 |
| $f_2$ | 129 | 5.31 | 5.72 | 191 |
| $f_2$' | 126 | 5.26 | 5.73 | 191 |
| $f_3$ | 2.3 | 3.3 | 11.4 | 460 |
| $f_4$ | 13 | 5.8 | 14 | 640 |
| $f_5$ | 27 | 12 | 21 | 700 |

Table 3. Parameters for the fit to water shown in Fig 7. β is the relaxation mode modeled as a Cole-Cole function and $f_1 - f_5$ are the Brownian oscillators (Eq. (11)) representing the librational modes. The fitted rise time parameter $\gamma_{LIB}$ = 12 THz. The parameters with a prime superscript refer to the alternative fit where the α termination is included. 10 THz is close to the upper limit of the spectrum and the parameters for the librational modes $f_3 - f_5$ are approximate.

## CONCLUSION

In addition to the structural α relaxation, the OKE data for NMA at temperatures above the melting point show a clear secondary β relaxation and a sub-picosecond temperature-independent decay. For the general case, rotational relaxation decays measured by four-wave-mixing experiments are given by five-exponential decays in the time domain. However, the presence of symmetry elements results in considerable simplifications, and one- or two-exponential decays are commonly measured. Non-Johari-Goldstein secondary relaxation modes can arise from intramolecular rotations and librations but, although NMA contains a single bonded C-N link in the backbone, it is well known that in the peptide linkage, rotation is severely restricted by delocalization of the (double) carbonyl bond, which renders the molecule planar. This is particularly true for methyl-substituted acetamides such as NMA, even at elevated temperatures.[41,63] The barrier to rotation of the methyl groups is relatively small but this mode is not significant here. Therefore, NMA is effectively a rigid molecule and, ignoring hydrogens, is planar with the trans conformation of the O=C-N-H moiety favored thermodynamically by approximately 10 kJ mol$^{-1}$ over the cis.[64] This is confirmed in the liquid phase at room temperature by spectroscopic studies of the amide-I band.[39] In similar amides, the polarizability ellipsoid is dominated by the double bond delocalized across the N-C-O moiety and is oriented with the major axis approximately perpendicular to the plane.[65,66] The molecule is then, to a first approximation, an oblate top. The lowest energy internal (deformation) modes are calculated to occur at above 1.5 THz,[66] so the observed secondary relaxation then involves the whole molecule and satisfies the fundamental Johari-Goldstein definition of β relaxation. However, without supercooling the liquid or studying the pressure dependence of the β relaxation, we cannot say if it satisfies the full criteria of a true Johari-Goldstein β relaxation, but the measured relaxation appears to be a genuine intermolecular process. An additional β relaxation at higher frequency is unresolved from the librational modes and cannot be accurately analyzed.

β relaxation is normally associated with glass-like behavior in the Johari-Goldstein model where it is related to the formation of transient cages in the (supercooled) liquid. Although, at these temperatures, the α-relaxation appears as a simple Debye function the β relaxation is well described by the Cole-Cole function. The Cole-Cole function was introduced as an empirical function to describe complex behavior but it has since been derived from fundamental models of diffusion and shown to be consistent with mode coupling theory as a model of β relaxation.[31,32] We employ it as a simple model which accurately describes the data and whose parameters can be related to the physical properties of the liquid with respect to its structure. The difference in the temperature dependence of the amplitude compared to that of the α relaxation is consistent with the presence of two separate relaxation modes.

The simple models derived from the Debye function are unphysical at short times. This shortcoming has previously been corrected by including a libration-dependent rise function in the time domain; we show how this modification can be introduced in the frequency domain in a quite general way.

We also consider the termination of a β relaxation by the α relaxation which must occur if, for example, the β relaxation is a cage-constrained mode and α relaxation is the complete decay of structural correlation involving the break-up of the cage. This process arises naturally in stochastic models of diffusion such as in the MCT framework; we show for the first time how it can be included in these simple analytical models.

Given the complexity of the OKE response for NMA and the similar behavior of the existing functions used to model relaxation in liquids, it is not possible to present a definitive model for the relaxation. Rather, a simple model of the intramolecular relaxation is applied here consisting of a β relaxation, which is an activated process such as a cage-constrained translational motion, which gives way to an α relaxation representing the lifetime of the cage.

For water; the measured relaxation is shown to be consistent with the higher frequency of the two modes measured by DRS. The relaxation measured in water by OKE is then seen to be β relaxation rather than (rotational) α relaxation, and can be fitted by a Cole-Cole function as can the β relaxation mode in NMA.

## APPENDIX

The Debye function is widely applied as a model of molecular relaxation but it is unphysical. In the time domain it does not start at a value of zero at $t = 0$, and therefore in the frequency domain it decays too slowly towards high frequency. Thus, in time domain OKE studies a modified Debye function has been used[11,16,53] which includes a rise function in the form

$$D'(t) = \theta(t) \frac{2\pi}{\tau} e^{-t/\tau} \left[1 - e^{-\gamma t}\right] \quad (16)$$

or alternatively

$$D'(t) = \theta(t) \frac{2\pi}{\tau} \left[e^{-t/\tau} - e^{-\gamma' t}\right] \quad (17)$$

which is identical to (16) except for a small parameter transformation in $\gamma$.

A frequency domain transformation of (17) has been presented but the equivalent for the Cole-Cole function was not found.[67] The analytical Fourier transform can be made for (16):

$$D'(\omega) = D(\omega) - D(\omega + i\gamma), \quad (18)$$

but not for the Cole-Cole function as no practical time domain form is available. Here we give a general solution for both the rise function and the α termination in the frequency domain. For a time-domain function $f(t)$, the Fourier transform is defined as

$$f(\omega) = (2\pi)^{-1} \int dt f(t) e^{+i\omega t} \quad (19)$$

and for a causal function, we can write

$$f(\omega) = (2\pi)^{-1} \int_{-\infty}^{\infty} dt \theta(t) f(t) e^{+i\omega t}, \quad (20)$$

where $\theta(t)$ is the Heaviside function, or equivalently,

$$f(\omega) = (2\pi)^{-1} \int_0^{\infty} dt f(t) e^{+i\omega t}. \quad (21)$$

To then modify the function by adding an α termination, we take the product $f(t) e^{-t/\tau_\alpha}$ hence,

$$\begin{aligned} f(\omega) &= (2\pi)^{-1} \int_0^{\infty} dt f(t) e^{-t/\tau_\alpha} e^{+i\omega t} \\ &= (2\pi)^{-1} \int_0^{\infty} dt f(t) e^{+i(\omega + i/\tau_\alpha)t} \quad (22) \\ &= f(\omega + i/\tau_\alpha) \end{aligned}$$

This expression is general and can be applied even to functions, such as the Cole-Cole, which do not have an analytical Fourier transform. Similarly, the inertial rise can be included by multiplying with $1 - e^{-\gamma_{LIB} t}$. The Fourier transform is

$$\begin{aligned} &F\left(f(t)\left[1 - e^{-\gamma_{LIB} t}\right] e^{-t/\tau_\alpha}\right) \\ &= F\left(f(t)\left[e^{-t/\tau_\alpha} - e^{-(\gamma_{LIB} + 1/\tau_\alpha)t}\right]\right) \quad (23) \\ &= f(\omega + i/\tau_\alpha) - f(\omega + i[\gamma_{LIB} + 1/\tau_\alpha]) \end{aligned}$$

Fig 8 shows the influence of both modifications on the Cole-Cole function.

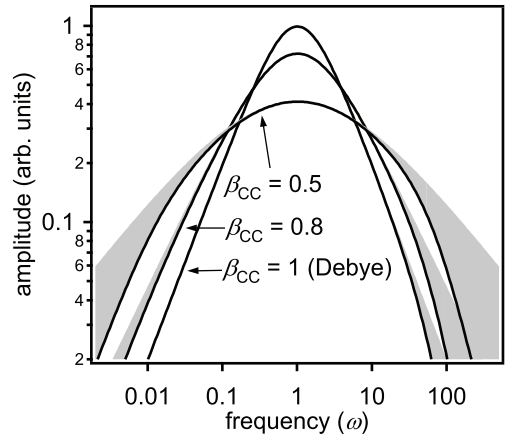

Fig 8. *The result of the modifications to the Cole-Cole function (Eq.(4)) by the rise function ( 16) and the α termination. The extent of the shaded areas represents the*

*unmodified functions with the Cole-Cole exponents of 1, 0.8, and 0.5. The frequency scale is normalized to $\tau = 1$. The solid lines are the modified functions for $\gamma = 20$ and $\tau_\alpha = 20$.*

## ACKNOWLEDGEMENTS

We gratefully acknowledge funding for this project from the Engineering and Physical Sciences Research Council (EPSRC) and the Leverhulme Trust.